# Twenty-Two New Variable Stars in the Northern Sky and Light Elements Improvement for PT Lyr, [WM2007] 1157, and [WM2007] 1160

**Riccardo Furgoni**
*Keyhole Observatory MPC K48, Via Fossamana 86, S. Giorgio di Mantova (MN), Italy; riccardo.furgoni@gmail.com*

*and*

*AAMN Gorgo Astronomical Observatory MPC 434, S. Benedetto Po (MN), Italy*

**Abstract**   I report the discovery of twenty-two new variable stars in the Northern Sky: eleven eclipsing, seven pulsating (one of which is suspected), three rotating, and one of unspecified type. Three already known variables (PT Lyr, [WM2007] 1157, and [WM2007] 1160) have been studied more carefully in order to clarify their characterization or provide updated ephemeris.

## 1. Introduction

A new photometric campaign aimed at the discovery of variable stars was carried out from the Keyhole Observatory (MPC K48) in S. Giorgio di Mantova, Italy. Three separate fields were observed, located in the constellations Cassiopeia, Lacerta, and both Lyra and Cygnus, over a period of forty-one nights, obtaining 4,492 images in the V passband with an overall exposure time of almost 150 hours. Subsequently, the light curves of all stars in the fields have been visually inspected in order to determine the candidate variables, starting with those that have an RMS scatter vs. magnitude higher than normal. When possible, the observations were combined with CTI-I Survey (Wetterer and McGraw 2007), NSVS (Woźniak *et al.* 2004), and SuperWASP (Butters *et al.* 2010) datasets in order to improve the precision in the determination of the period, magnitude range, and the type of variability.

Some known variables in the fields (PT Lyr, [WM2007] 1157, and [WM2007] 1160) have been studied more carefully in order to clarify their characterization or provide updated ephemeris.

## 2. Instrumentation used

The data were obtained with a Celestron C8 Starbright, a Schmidt-Cassegrain optical configuration with aperture of 203 mm and central obstruction of 34%.

The telescope was positioned at coordinates 45°12'33"N 10°50'20"E (WGS84) at the Keyhole Observatory, a roll-off roof structure managed by the author. The telescope was equipped with a focal reducer Baader Planetarium Alan Gee II able to bring the focal length from 2030 mm to 1413 mm in the optical train used. The focal ratio is down to f/6.96 from the original f/10.

The pointing was maintained with a Syntha NEQ6 mount guided via a Baader Vario Finder telescope equipped with a Barlow lens capable of bringing the focal length of the system to 636 mm and focal ratio of f/10.5. The guide camera used was a Magzero MZ-5 with a Micron MT9M001 monochrome sensor. The CCD camera used was a SBIG ST8300m with monochrome sensor Kodak KAF8300. Photometry in the Johnson V passband was performed with an Astrodon Photometrics Johnson-V 50 mm round unmounted filter, equipping a Starlight Xpress USB filterwheel.

The camera is equipped with a 1000X antiblooming: after exhaustive testing it has been verified that the zone of linear response is between 1000 and 20000 ADU, although up to 60000 ADU the loss of linearity is less than 5%. The CCD is equipped with a single-stage Peltier cell $\Delta T = 35 \pm 0.1°C$ which allows the cooling at a stationary temperature.

### 3. Data collection

The observed fields are centered respectively at coordinates (J2000) R.A. $19^h 19^m 00^s$, Dec. $+28° 03' 30''$, R.A. $22^h 41^m 00^s$, Dec. $+48° 00' 00''$, and R.A. $02^h 53^m 09^s$, Dec. $+62° 02' 00''$. The dimensions for all are $44' \times 33'$ with a position angle of $360°$.

These fields were chosen to maximize the possibility of discovering new variable stars. Their determination was made trying to meet the following criteria:

• low galactic latitude to have a greater number of sources;

• equatorial coordinates compatible with low air masses for most of the night in the observing site;

• low or no presence of already known variables, where this element has been determined by analyzing the distribution of the variables in the International Variable Star Index (Watson *et al.* 2007) operated by the AAVSO.

The observations were performed with the CCD at a temperature of –10°C (in summer and fall) and –20°C (in winter) in binning $1 \times 1$. The exposure time was 120 seconds with a delay of 1 second between the images and an average download time of 11 seconds per frame. Once the images were obtained, the calibration frames were taken for a total of 100 dark of 120 seconds each, 200 darkflat of 2 seconds, and 50 flat of 2 seconds. The darkflats and darks were taken only the first observation session for every field and used for all other sessions. The flats were taken for each session as the position of the CCD camera could be varied slightly, as well as the focus point.

The calibration frames were combined with the method of the median, and the masterframes obtained were then used for the correction of the images taken. All images were then aligned and an astrometric reduction was made to implement the astrometrical coordinate system WCS in the FITS header. These operations were conducted entirely through the use of MAXIMDL version 5.23 (Diffraction Limited 2012).

## 4. Methods and procedures

This work is a continuation of the wider survey activity conducted from the Keyhole Observatory-MPC K48 for the discovery of new variable stars. In this section, the methods and procedures used are presented in a very schematic way; details can be found in the author's previously published work (see, in particular, Furgoni (2013a and 2013b)). The technique used for the determination of the magnitude was differential photometry, ensemble-type, in almost all cases. As is known, it is necessary to have comparison stars with magnitudes accurately determined in order to obtain reliable measurements, and as close to the standard system as possible. Since the entire observational campaign was conducted with the use of the Johnson V filter, the comparison star magnitudes were obtained:

• as a first choice from the APASS (Henden *et al.* 2013) V-magnitude data provided by the AAVSO;

• as a second choice deriving the V magnitude from the CMC14 (Copenhagen Univ. Obs. 2006) r' magnitude as described in Dymock and Miles (2009).

In both cases the calculated magnitude for the check star was only slightly different from the respective APASS V magnitude and CMC14-derived V magnitude. In fact, in the first case, the average error was normally of the order of 0.01–0.02 magnitude, while in the second case it never exceeded 0.05 magnitude. These errors are stable even when using stars with moderately different color indices.

The search for new variable stars was performed using the same technique described in the author's previously published work; the method involves the distinction between discovery-nights and follow-up nights. The first is needed to determine if the chosen field contains potential candidate variables through the creation of a magnitude-RMS diagram and the accurate visual inspection of all light curves down to sixteenth magnitude. If in the discovery-night the chosen field does not appear interesting, it is discarded without further analysis. The follow-up nights are necessary for the collection of data concerning variable star candidates. In any case, any follow-up night is checked again by a magnitude-RMS scatter diagram. In this case, however, the light curves of the stars in the field are not inspected visually due to the long time needed for the operation. The combination of these two methods ensures a good compromise between the need to maximize the chance of discovery and the need to obtain the maximum from the data gradually collected.

Before proceeding further with the analysis, the time of the light curves obtained was heliocentrically corrected (HJD) in order to ensure perfect compatibility of the data with other observations carried out, even at a considerable distance in time. When necessary, the determination of the period was calculated using the software PERIOD04 (Lenz and Breger 2005), using a Discrete Fourier Transform (DFT). The average zero-point (average magnitude of the object) was subtracted from the dataset to prevent the appearance of artifacts centered at a frequency 0.0 of the periodogram. The calculation of the uncertainties was carried out with PERIOD04 using the method described in Breger *et al.* (1999).

To improve the period determination, the NSVS and SWASP photometric data were used when available. However, due to high scattering which in some cases affects them, the data with high uncertainties were eliminated. The NSVS and SWASP data were also corrected in their zero-point to make them compatible with the author's V-band data. Having the same zero-point is indeed crucial for correct calculation of the PERIOD04 Discrete Fourier Transform.

## 5. New variable stars and known variables studied

In this survey twenty-two new variable stars were discovered and three already known variables (PT Lyr, [WM2007] 1157, and [WM2007] 1160) were studied more carefully. The population of the new variables is as follows:

- 11 eclipsing (4 βLyr, 4 β Per, and 3 W UMa)
- 7 pulsating (6 δ Sct (one suspected), 1 Semiregular)
- 3 rotating (3 rotating ellipsoidal)
- 1 unspecified (but with evident signs of periodicity)

The coordinates of all new variable stars discovered in this survey are reported as they appear in the UCAC4 catalogue (Zacharias *et al.* 2012) and differ from the detected positions by never more than 0.5".

### 5.1. [WM2007] 1157

Position (UCAC4): R.A. (J2000) = $19^h 18^m 27.56^s$, Dec. (J2000) = +28° 01' 55.0"

Cross Identification: 2MASS J19182756+2801550; UCAC4 591-078527

Variability Type: δ Sct

Magnitude: Max. 14.30 V, Min. 14.49 V

Main Period: 0.043525536(4) d

Secondary Period: 0.05312469(1) d

Epoch Main Period: 2456505.39440(27) HJD

Epoch Secondary Period: 2456505.54615(54) HJD

Ensemble Comparison Stars: UCAC4 591-078539 (CMC14 derived 13.542 V); UCAC4 591-078567 (CMC14 derived 13.890 V)

Check Star: UCAC4 591-078567

Notes: $J–K = 0.078$. The star shows a clear, continuous modulation of the light curve.

Fourier spectrum and phase plots are shown in Figures 1, 2, and 3.

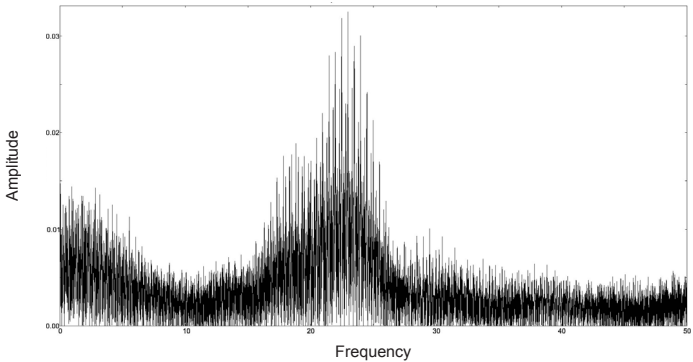

Figure 1. Fourier spectrum of [WM2007] 1157.

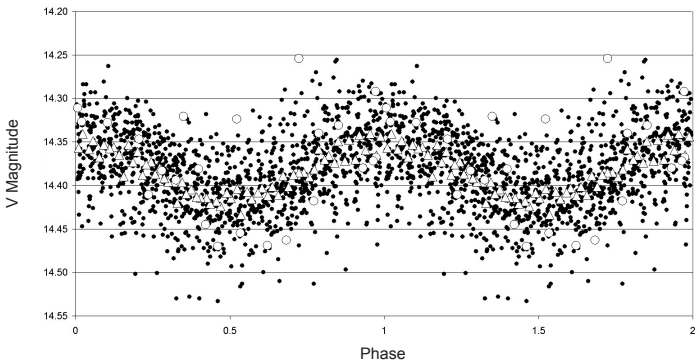

Figure 2. Main Period Phase plot of [WM2007] 1157. Filled circles denote Furgoni data; open triangles denote Furgoni data (binning 20); open circles denote CTI-I survey data.

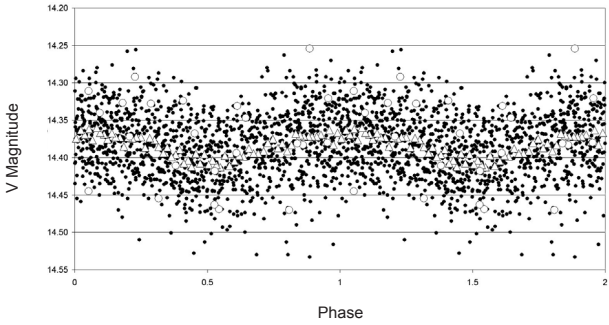

Figure 3. Secondary Period Phase plot of [WM2007] 1157. Filled circles denote Furgoni data; open triangles denote Furgoni data (binning 20); open circles denote CTI-I survey data.

5.2. [WM2007] 1160

Position (UCAC4): R.A. (J2000) = 19$^h$ 20$^m$ 03.51$^s$, Dec. (J2000) = +28° 04' 24.4"

Cross Identification: 2MASS J19200351+2804246; UCAC4 591-079175

Variability Type: W UMa

Magnitude: Max. 15.50 V, Min. 16.08 V

Period: 0.4978081(1) d

Epoch: 2456462.5052(7) HJD

Ensemble Comparison Stars: UCAC4 591-079193 (CMC14 derived 14.326 V); UCAC4 591-079139 (CMC14 derived 13.806 V)

Check Star: UCAC4 591-079214

Notes: J–K = 0.45. Phase Plot is shown in Figure 4.

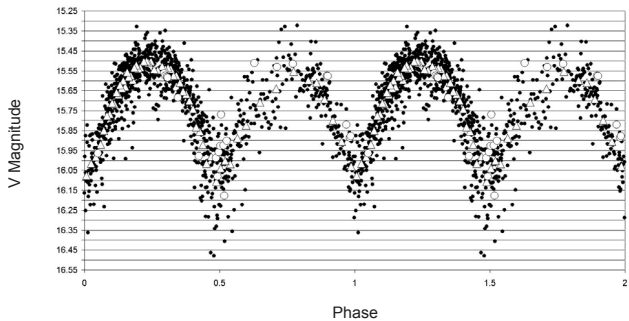

Figure 4. Phase plot of [WM2007] 1160. Filled circles denote Furgoni data; open triangles denote Furgoni data (binning 20); open circles denote CTI-I survey data.

5.3. PT Lyr

Position (UCAC4): R.A. (J2000) = 19$^h$ 18$^m$ 29.40$^s$, Dec. (J2000) = +27° 55′ 41.4″

Cross Identification: VV 108

Variability Type: RR Lyrae (ab type)

Magnitude: Max. 15.41 V, Min, 16.80 V

Period: 0.5159309(15) d

Epoch: 2456462.515(2) HJD

Ensemble Comparison Stars: UCAC4 590-080143 (CMC14 derived 13.298 V); UCAC4 590-080184 (CMC14 derived 14.295 V)

Check Star: UCAC4 590-080067

Notes: Rise duration 13%. Phase plot is shown in Figure 5.

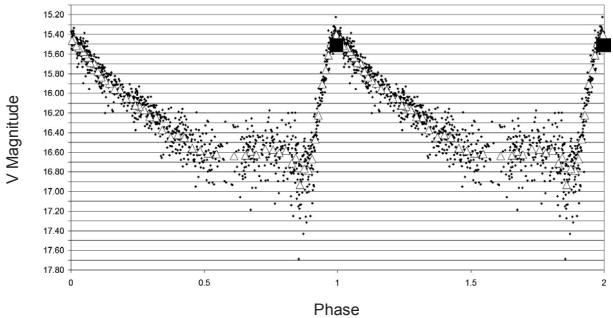

Figure 5. Phase plot of PT Lyr. Filled circles denote Furgoni data; open triangles denote Furgoni data (binning 20); large squares denote VV108 published time of maximum (Miller and Wachmann 1963).

5.4. 2MASS J19181000+2801294

Position (UCAC4): R.A. (J2000) = 19$^h$ 18$^m$ 10.00$^s$, Dec. (J2000) = 28° 01′ 29.6″

Cross Identification: UCAC4 591-078397

Variability Type: β Per

Magnitude: Max. 15.83 V, Min. 16.82 V

Period: 0.89394(12) d

Epoch: 2456508.3230(42) HJD

Ensemble Comparison Stars: UCAC4 591-078539 (CMC14 derived 13.542V); UCAC4 591-078667 (CMC14 derived 13.676 V)

Check Star: UCAC4 591-078567

Notes: J–K = 0.23. Eclipse duration 16%. Phase plot is shown in Figure 6.

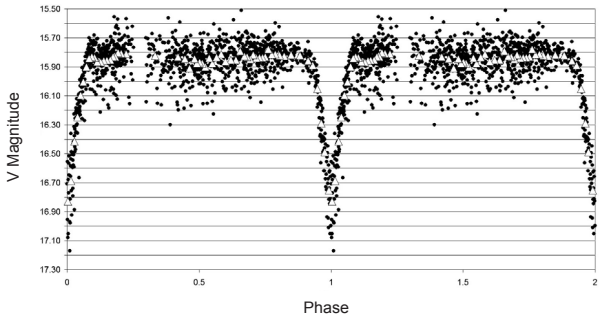

Figure 6. Phase plot of 2MASS J19181000+2801294. Filled circles denote Furgoni data; open triangles denote Furgoni data (binning 20).

### 5.5. 2MASS J19194398+2758145

Position (UCAC4): R.A. (J2000) = 19$^h$ 19$^m$ 43.98$^s$, Dec. (J2000) = 27° 58' 14.5"

Cross Identification: NOMAD1 1179-0436600

Variability Type: W UMa

Magnitude: Max. 15.6 V, Min. 16.4 V

Period: 0.311487(7) d

Epoch: 2456462.5530(16) HJD

Ensemble Comparison Stars: UCAC4 590-080783 (CMC14 derived 11.119 V); UCAC4 590-080567 (CMC14 derived 11.781 V)

Check Star: UCAC4 590-080651

Notes: Aperture photometry was performed on the blended source formed by the variable and the close companion 2MASS J19194382+2758185 that lies 4.4" away. Subsequently, the magnitude of the companion (13.45 V) was measured from a stack of 30 frames (in order to improve the SNR and highlight the small separation) and subtracted from the blended source. Phase plot is shown in Figure 7.

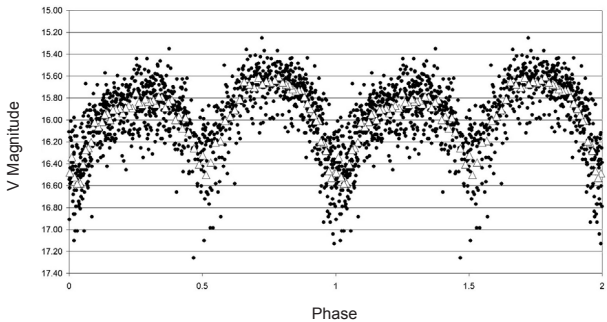

Figure 7. Phase plot of 2MASS J19194398+2758145. Filled circles denote Furgoni data; open triangles denote Furgoni data (binning 15).

5.6. GSC 02132-03510

Position (UCAC4): R.A. (J2000) = 19$^h$ 17$^m$ 52.22$^s$, Dec. (J2000) = +27° 54' 31.7"

Cross Identification: 2MASS J19175221+2754317; UCAC4 590-079883

Variability Type: δ Sct

Magnitude: Max. 11.62 V, Min. 11.64 V

Period: 0.096578(4) d

Epoch: 2456495.41908(7) HJD

Ensemble Comparison Stars: UCAC4 590-080116 (CMC14 derived 11.275 V); UCAC4 590-080091 (CMC14 derived 12.223 V)

Check Star: UCAC4 590-080046

Notes: J–K = 0.23. Phase plot is shown in Figure 8.

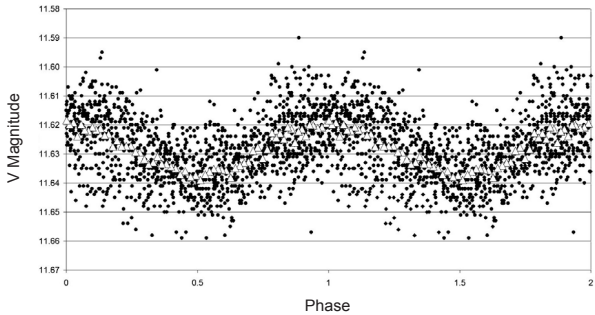

Figure 8. Phase plot of GSC 02132-03510. Filled circles denote Furgoni data; open triangles denote Furgoni data (binning 20).

5.7. GSC 02136-03136

Position (UCAC4): R.A. (J2000) = 19$^h$ 20$^m$ 34.93$^s$, Dec. (J2000) = +28° 13' 31.6"

Cross Identification: 2MASS J19203492+2813315; UCAC4 592-078257

Variability Type: δ Sct

Magnitude: Max. 12.475 V, Min. 12.495 V

Period: 0.079901(3) d

Epoch: 2456505.5814(7) HJD

Ensemble Comparison Stars: UCAC4 591-078993 (CMC14 derived 10.762 V); UCAC4 591-079007 (CMC14 derived 12.075 V)

Check Star: UCAC4 591-078915

Notes: J–K = 0.23. Phase plot is shown in Figure 9.

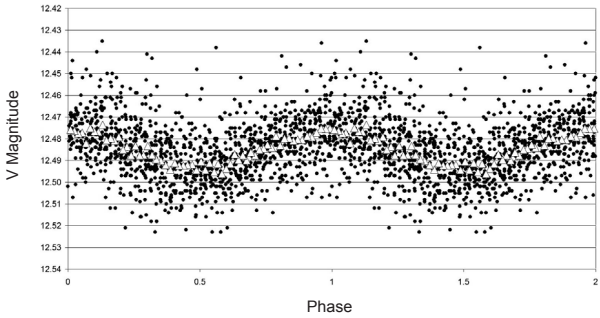

Figure 9. Phase plot of GSC 02136-03136. Filled circles denote Furgoni data; open triangles denote Furgoni data (binning 20).

### 5.8. GSC 03625-01173

Position (UCAC4): R.A. (J2000) = 22$^h$ 42$^m$ 09.90$^s$, Dec. (J2000) +47° 45' 26.7"

Cross Identification: 2MASS J22420989+4745267; UCAC4 689-119504; NSVS 6068811

Variability Type: β Lyr

Magnitude: Max 12.57 V, Min. 12.95 V

Period: 0.9921132(3) d

Epoch: 2456560.418(2) HJD

Ensemble Comparison Stars: UCAC4 689-119409 (APASS 12.382 V); UCAC4 689-119403 (APASS 12.443 V)

Check Star: UCAC4 689-119442

Notes: Secondary minimum = 12.80 V. Phase plot is shown in Figure 10.

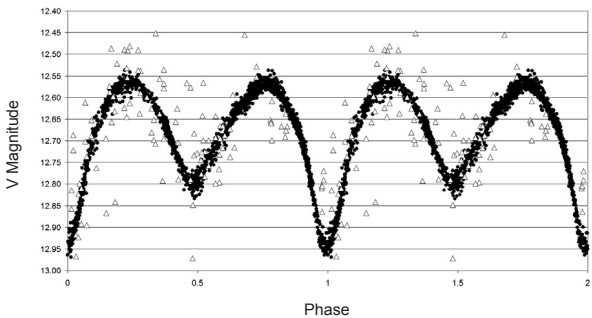

Figure 10. Phase plot of GSC 03625-01173. Filled circles denote Furgoni data; open triangles denote NSVS data (with error less than 0.1 magnitude).

5.9. 2MASS J22405571+4804277

Position (UCAC4): R.A. (J2000) = 22ʰ 40ᵐ 55.72ˢ, Dec. (J2000) = +48° 04' 27.9"

Cross Identification: UCAC4 691-117787; NSVS 6067528

Variability Type: β Lyr

Magnitude: Max. 13.57 V, Min. 14.16 V

Period: 0.763656(16) d

Epoch: 2456559.3958(12) HJD

Ensemble Comparison Stars: UCAC4 691-117919 (CMC14 derived 12.358 V); UCAC4 691-117825 (CMC14 derived 12.301 V)

Check Star: UCAC4 691-117831

Notes: Secondary minimum = 13.75 V. Phase plot is shown in Figure 11.

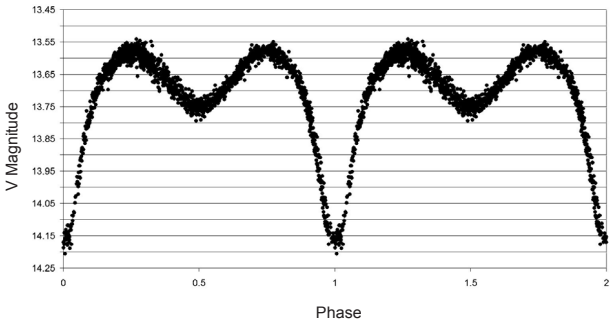

Figure 11. Phase plot of 2MASS J22405571+4804277. Filled circles denote Furgoni data.

5.10. 2MASS J22421092+4807300

Position (UCAC4): R.A. (J2000) = 22ʰ 42ᵐ 10.93ˢ, Dec. (J2000) = +48° 07' 30.0"

Cross Identification: UCAC4 691-118072; USNO-B1.0 1381-0557121

Variability Type: W UMa

Magnitude: Max. 15.50 V, Min. 15.67 V

Period: 0.267008(3) d

Epoch: 2456559.404(1) HJD

Ensemble Comparison Stars: UCAC4 691-118056 (CMC14 derived 14.835 V); UCAC4 691-118045 (CMC14 derived 14.969 V)

Check Star: UCAC4 691-118048

Notes: Phase plot is shown in Figure 12.

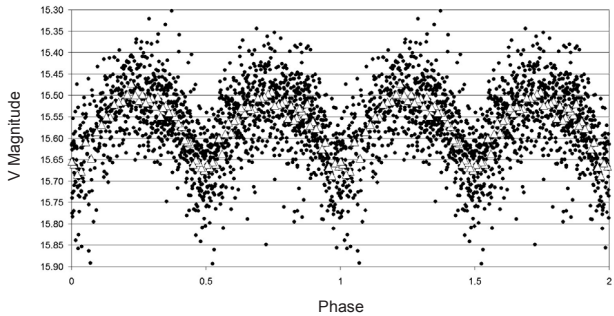

Figure 12. Phase plot of 2MASS J22421092+4807300. Filled circles denote Furgoni data; open triangles denote Furgoni data (binning 20).

5.11. GSC 03624-02592

Position (UCAC4): R.A. (J2000) = 22$^h$ 40$^m$ 30.81$^s$, Dec. (J2000) +47° 48' 10.5"

Cross Identification: 2MASS J22403080+4748105; UCAC4 690-119955; NSVS 6067049

Variability Type: δ Sct

Magnitude: Max .12.79 V, Min. 12.82 V

Period: 0.072472(2) d

Epoch: 2456549.4393(7) HJD

Ensemble Comparison Stars: UCAC4 690-119902 (CMC14 derived 11.383 V); UCAC4 690-119823 (CMC14 derived 11.625 V)

Check Star: UCAC4 690-120037
Notes: Phase plot is shown in Figure 13.

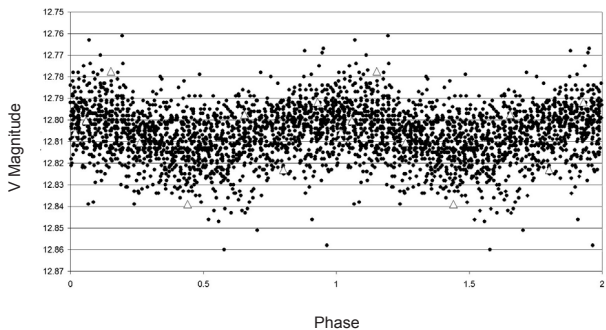

Figure 13. Phase plot of GSC 03624-02592. Filled circles denote Furgoni data; open triangles denote NSVS data (error less than 0.1 magnitude; binning 20).

5.12. GSC 03624-02115

Position (UCAC4): R.A. (J2000) = 22ʰ 40ᵐ 57.49ˢ, Dec. (J2000) = 47° 54' 20.9"

Cross Identification: 2MASS J22405748+4754206; UCAC4 690-120057; NSVS 6067581

Variability Type: Semiregular Late-type (SRB)

Magnitude: Max. 12.65 V, Min. 13.02 V

Period: 55.5(1) d

Epoch: 2456531.7(5) HJD

Ensemble Comparison Stars: UCAC4 690-119902 (CMC14 derived 11.383 V); UCAC4 690-119823 (CMC14 derived 11.625 V)

Check Star: UCAC4 690-120037

Notes: J–K = 1.25. NSVS magnitude contaminated by GSC 03624-02481 (V = 12.1; sep. 46") and other nearby stars. NSVS range has been corrected and shifted in the phase plot. Light curve and phase plot are shown in Figures 14 and 15.

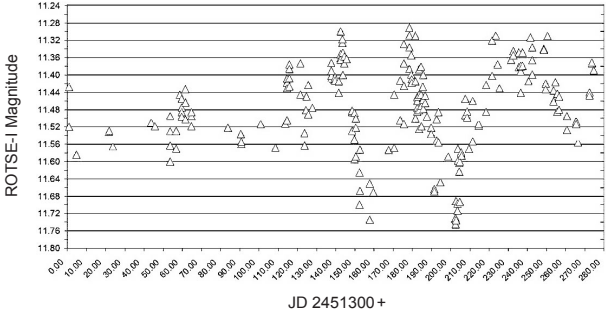

Figure 14. NSVS light curves of GSC 03624-02115 (after subtraction of the blended source GSC 03624-02481; V 12.1; separation 46").

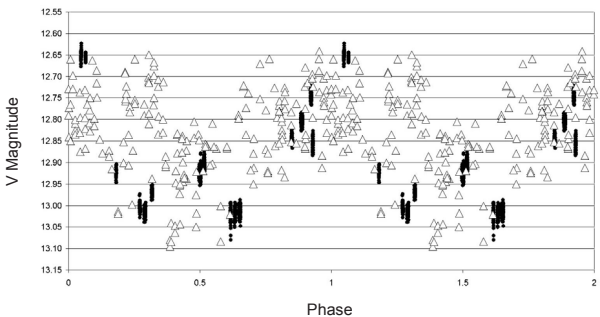

Figure 15. Phase plot of GSC 03624-02115. Filled circles denote Furgoni data; open triangles denote NSVS data (magnitude of the blended source GSC 03624–02481; +1.35 magnitude offset).

5.13. GSC 03625-01798

Position (UCAC4): R.A. (J2000) = 22$^h$ 42$^m$ 59.93$^s$, Dec. (J2000) = +47° 57' 43.4"

Cross Identification: 2MASS J22425993+4757434; UCAC4 690-120491

Variability Type: Rotating ellipsoidal

Magnitude: Max. 13.79 V, Min. 13.85 V

Period: 0.55392(2) d

Epoch: 2456623.323(2) HJD

Ensemble Comparison Stars: UCAC4 690-120342 (CMC14 derived 11.762 V); UCAC4 690-120233 (CMC14 derived 11.810 V)

Check Star: UCAC4 690-120294

Notes: J–K = 0.65. Secondary minimum = 13.83 V. Phase plot is shown in Figure 16.

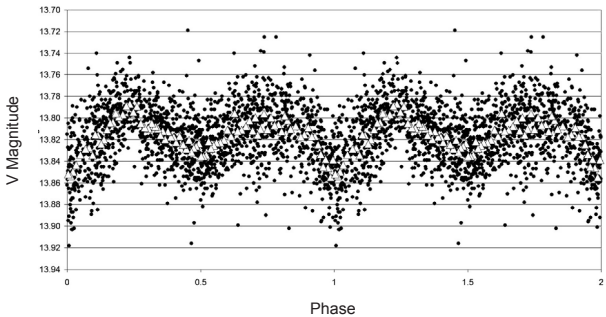

Figure 16. Phase plot of GSC 03625-01798. Filled circles denote Furgoni data; open triangles denote Furgoni data (binning 20).

5.14. GSC 04052-01292

Position (UCAC4): R.A. (J2000) = 02$^h$ 52$^m$ 44.53$^s$, Dec. (J2000) = +62° 00' 13.4"

Cross Identification: 2MASS J02524451+6200133; UCAC4 761-021866; NSVS 1888262

Variability Type: β Per

Magnitude: Max. 13.88 V, Min. 14.62 V

Period: 1.00599(8) d

Epoch: 2456712.241(1) HJD

Ensemble Comparison Stars: UCAC4 760-023030 (APASS 11.963 V); UCAC4 760-022993 (APASS 11.608 V)

Check Star: UCAC4 760-022973

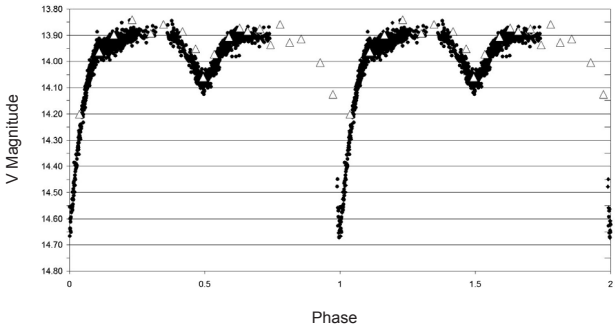

Figure 17. Phase plot of GSC 04052-01292. Filled circles denote Furgoni data; open triangles denote NSVS data (+0.16 magnitude offset; binning 20).

Notes: Eclipse duration 20%. Secondary minimum = 14.07 V. Phase plot is shown in Figure 17.

5.15. GSC 04051-02709

Position (UCAC4): R.A. (J2000) = 02$^h$ 50$^m$ 12.66$^s$, Dec. (J2000) = 62° 04' 55.0"

Cross Identification: 2MASS J02501265+6204549; UCAC4 761-021561

Variability Type: β Per

Magnitude: Max. 12.88 V, Min. 13.22 V

Period: 1.57580(6) d

Epoch: 2456711.250(4) HJD

Ensemble Comparison Stars: UCAC4 761-021680 (APASS 12.068 V); UCAC4 761-021669 (APASS 12.142 V)

Check Star: UCAC4 761-021632

Notes: Eclipse duration 18%. Secondary minimum = 13.15 V. Phase plot is shown in Figure 18.

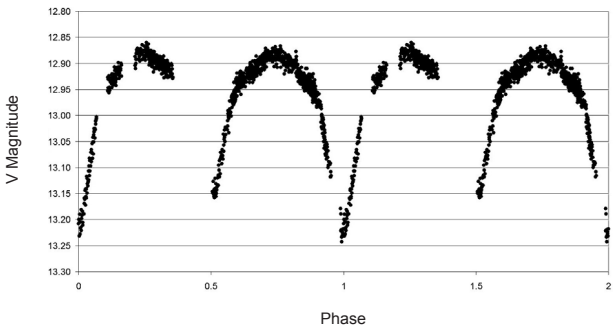

Figure 18. Phase plot of GSC 04051-02709. Filled circles denote Furgoni data..

5.16. GSC 04052-01378

Position (UCAC4): R.A. (J2000) = $02^h 53^m 08.34^s$, Dec. (J2000) = +62° 06' 10.5"

Cross Identification: UCAC4 761-021922; NSVS 1888562; 1SWASP J025308.36+620610.7

Variability Type: β Per

Magnitude: Max. 11.76 V, Min. 12.08: V

Spectral type: B2

Period: 18.3024(1) d

Epoch: 2451403.83(1) HJD

Ensemble Comparison Stars: UCAC4 761-021906 (APASS 12.498 V); UCAC4 761-021905 (APASS 12.566 V)

Check Star: UCAC4 761-022036

Notes: Primary eclipse duration 2%:. Secondary minimum = 12.06 V with eclipse duration 6%:. The great difference between eclipse duration in Min. and Min. II is indicative of a very eccentric system. Min. II observations are better phased with slightly shorter period suggesting possible apsidal motion. Elements for Min. II are HJD 2451486.75 + 18.3007 × E. Spectral type taken from Skiff (2009–2013). Phase plots are shown in Figures 19, 20, 21, and 22.

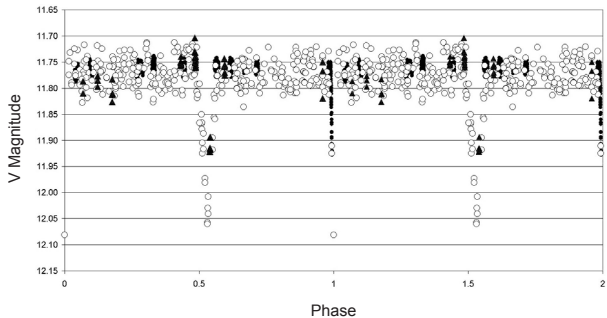

Figure 19. Phase plot of GSC 04052-01378. Filled circles denote Furgoni data; open circles denote NSVS data (+0.15 magnitude offset); filled triangles denote SWASP data (+0.08 magnitude offset).

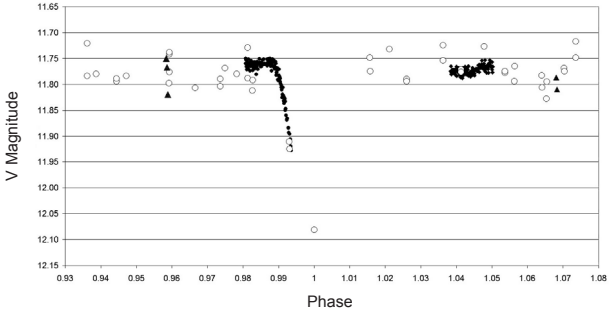

Figure 20. Phase plot of GSC 04052-01378 (Zoom on primary minimum). Filled circles denote Furgoni data; open circles denote NSVS data (+0.15 magnitude offset); filled triangles denote SWASP data (+0.08 magnitude offset).

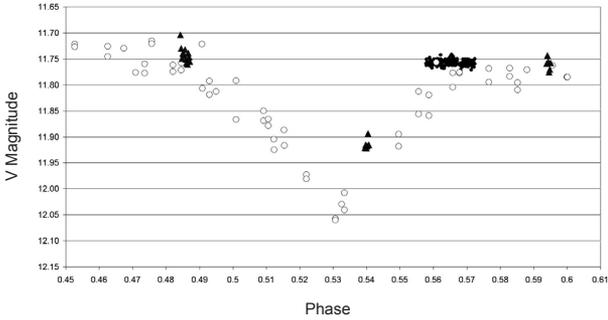

Figure 21. Phase plot of GSC 04052-01378 (Zoom on secondary minimum). Filled circles denote Furgoni data; open circles denote NSVS data (+0.15 magnitude offset); filled triangles denote SWASP data (+0.08 magnitude offset).

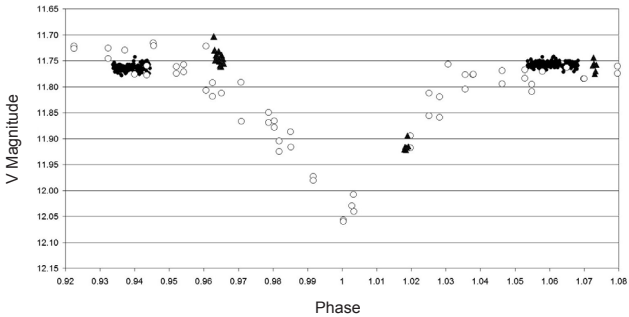

Figure 22. Phase plot of GSC 04052-01378 (Zoom on secondary minimum phased on its own elements; possible apsidal motion). Filled circles denote Furgoni data; open circles denote NSVS data (+0.15 magnitude offset); filled triangles denote SWASP data (+0.08 magnitude offset). Period = 18.3007 d; HJDmin = 2451486.75.

5.17. 2MASS J02530428+6208007

Position (UCAC4): R.A. (J2000) = 02$^h$ 53$^m$ 04.31$^s$, Dec. (J2000) = +62° 08' 00.8"

Cross Identification: UCAC4 761-021911; USNO-B1.0 1521-0094336

Variability Type: β Lyr

Magnitude: Max. 14.06 V, Min. 14.23 V

Period: 0.81069(2) d

Epoch: 2456631.343(1) HJD

Ensemble Comparison Stars: UCAC4 761-021906 (APASS 12.498 V);
UCAC4 761-021905 (APASS 12.566 V)

Check Star: UCAC4 761-022036

Notes: Secondary minimum = 14.17 V and secondary maximum = 14.10 V. Phase plot is shown in Figure 23.

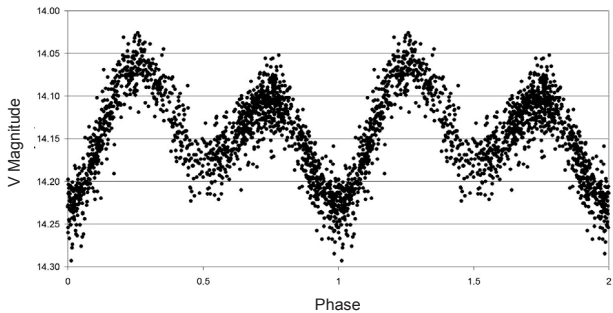

Figure 23. Phase plot of 2MASS J02530428+6208007. Filled circles denote Furgoni data.

5.18. 2MASS J02524261+6157132

Position (UCAC4): R.A. (J2000) = 02$^h$ 52$^m$ 42.65$^s$, Dec. (J2000) = +61° 57' 13.2"

Cross Identification: UCAC4 760-023072; 1SWASP J025242.72+615713.1

Variability Type: β Lyr

Magnitude: Max. 14.90 V, Min. 15.27 V

Period: 1.56250(9) d

Epoch: 2456713.405(3) HJD

Ensemble Comparison Stars: UCAC4 760-023030 (APASS 11.963 V);
UCAC4 760-022993 (APASS 11.608 V)

Check Star: UCAC4 760-022973

Notes: Secondary minimum = 15.05 V. Phase plot is shown in Figure 24.

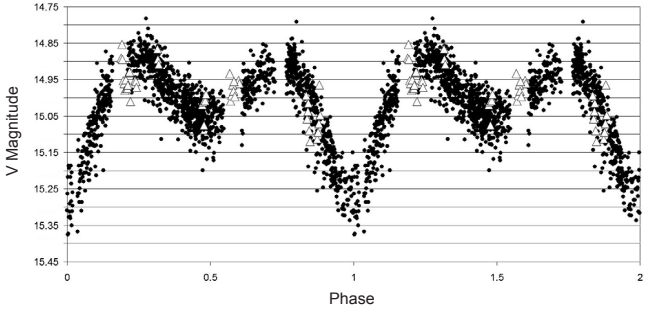

Figure 24. Phase plot of 2MASS J02524261+6157132. Filled circles denote Furgoni data; open triangles denote SWASP data (+1.53 magnitude offset).

### 5.19. GSC 04048-00893

Position (UCAC4): R.A. (J2000) = 02$^h$ 51$^m$ 29.89$^s$, Dec. (J2000) = +61° 47' 11.3"

Cross Identification: 2MASS J02512989+6147113; UCAC4 759-022788; 1SWASP J025129.87+614711.4

Variability Type: W UMa

Magnitude: Max. 13.82 V, Min. 14.31 V

Period: 0.433010(2) d

Epoch: 2456711.3693(5) HJD

Ensemble Comparison Stars: UCAC4 760-022986 (APASS 12.842 V); UCAC4 759-022881 (APASS 13.308 V)

Check Star: UCAC4 760-022963

Notes: Possible presence of O'Connell effect with secondary maximum = 13.84 V. Phase plot is shown in Figure 25.

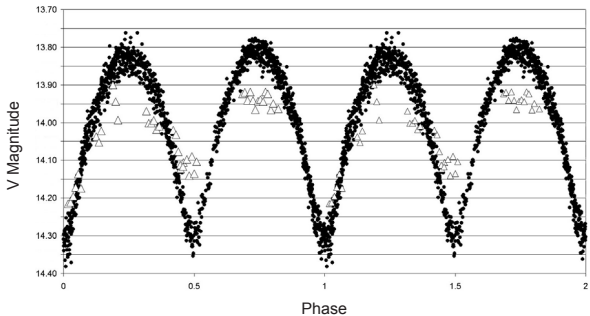

Figure 25. Phase plot of GSC 04048-00893. Filled circles denote Furgoni data; open triangles denote SWASP data (+0.95 magnitude offset).

5.20. GSC 04052-01674

Position (UCAC4): R.A. (J2000) = 02$^h$ 55$^m$ 39.74$^s$, Dec. (J2000) = +62° 10' 08.5"

Cross Identification: 2MASS J02553971+6210083; UCAC4 761-022258

Variability Type: δ Sct

Magnitude: Max. 12.855 V, Min. 12.875 V

Period: 0.0390453(4) d

Epoch: 2456698.3064(3) HJD

Ensemble Comparison Stars: UCAC4 761-022128 (APASS 12.013 V); UCAC4 761-022146 (APASS 12.741 V)

Check Star: UCAC4 762-023189

Notes: Phase plot is shown in Figure 26.

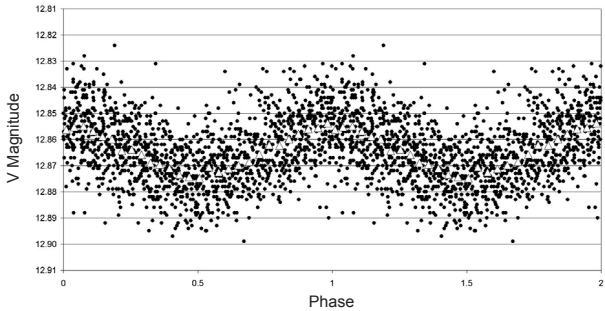

Figure 26. Phase plot of GSC 04052-01674. Filled circles denote Furgoni data; open triangles denote Furgoni data (binning 50).

5.21. GSC 04052-01198

Position (UCAC4): R.A. (J2000) = 02$^h$ 54$^m$ 22.56$^s$, Dec. (J2000) = +62° 13' 27.2"

Cross Identification: 2MASS J02542256+6213271; UCAC4 762-023119

Variability Type: δ Sct

Magnitude: Max. 12.06 V, Min. 12.07 V

Period: 0.072480(2) d

Epoch: 2456698.3438(7) HJD

Ensemble Comparison Stars: UCAC4 761-022128 (APASS 12.013 V); UCAC4 761-022146 (APASS 12.741 V)

Check Star: UCAC4 762-023189

Notes: Phase plot is shown in Figure 27.

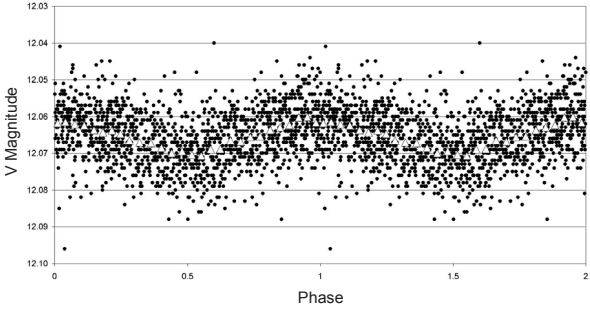

Figure 27. Phase plot of GSC 04052-01198. Filled circles denote Furgoni data; open triangles denote Furgoni data (binning 50).

### 5.22. GSC 04052-01238

Position (UCAC4): R.A. (J2000) 02$^h$52$^m$26.85$^s$, Dec. (J2000) = +61° 57' 00.4"

Cross Identification: 2MASS J02522686+6157004; UCAC4 760-023032

Variability Type: δ Sct (uncertain)

Magnitude: Max. 13.21 V, Min. 13.23 V

Period: 0.124125(6) d

Epoch: 2456630.470(2) HJD

Ensemble Comparison Stars: UCAC4 760-023030 (APASS 11.963 V); UCAC4 760-022993 (APASS 11.608 V)

Check Star: UCAC4 760-022973

Notes: B–V = 0.80; J–K = 0.40. Probably a reddened δ Sct but it might be short period ELL. Phase plot is shown in Figure 28.

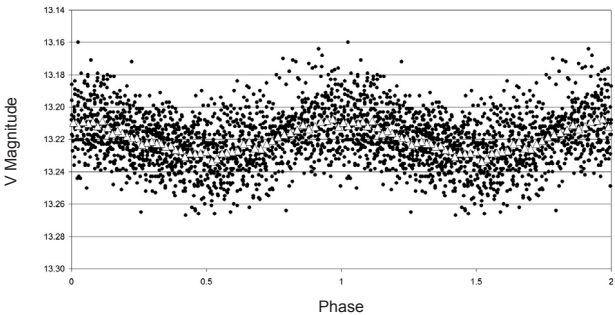

Figure 28. Phase plot of GSC 04052-01238. Filled circles denote Furgoni data; open triangles denote Furgoni data (binning 30).

5.23. 2MASS J02533682+6153083

Position (UCAC4): R.A. (J2000) = 02$^h$ 53$^m$ 36.81$^s$, Dec. (J2000) = +61° 53' 08.4"

Cross Identification: UCAC4 760-023199; USNO-B1.0 1518-0086484

Variability Type: Rotating ellipsoidal

Magnitude: Max. 14.48 V, Min. 14.55 V

Period: 0.58555(2) d

Epoch: 2456631.421(1) HJD

Ensemble Comparison Stars: UCAC4 760-023181 (APASS 11.403 V); UCAC4 760-023174 (APASS 11.373 V)

Check Star: UCAC4 760-023206

Notes: Phase plot is shown in Figure 29.

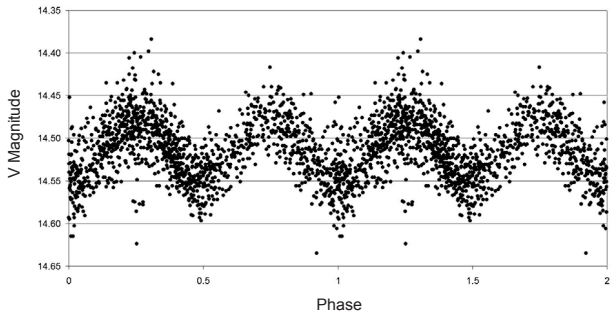

Figure 29. Phase plot of 2MASS J02533682+6153083. Filled circles denote Furgoni data.

5.24. GSC 04052-00946

Position (UCAC4): R.A. (J2000) = 02$^h$ 54$^m$ 59.05$^s$, Dec. (J2000) = +62° 03' 11.2"

Cross Identification:  J02545903+6203112; UCAC4 760-023199

Variability Type: Rotating ellipsoidal

Magnitude: Max. 11.43 V, Min. 11.46 V

Spectral Type: A5V

Period: 1.10349(4) d

Epoch: 2456701.462(4) HJD

Ensemble Comparison Stars: UCAC4 761-022213 (APASS 11.238 V); UCAC4 761-022128 (APASS 12.013 V)

Check Star: UCAC4 761-022184

Notes: Spectral type taken from Skiff (2009–2013). Phase plot is shown in Figure 30.

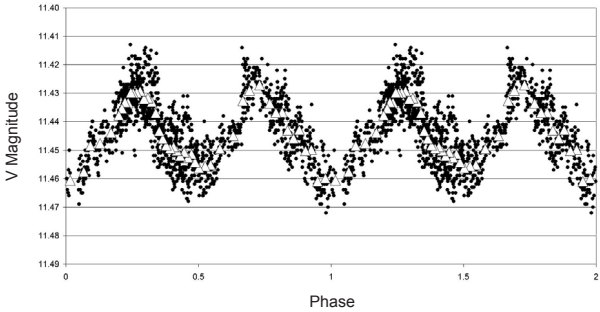

Figure 30. Phase plot of GSC 04052-00946. Filled circles denote Furgoni data; open triangles denote Furgoni data (binning 30).

5.25. GSC 04048-00441

Position (UCAC4): R.A. (J2000) = 02$^h$ 53$^m$ 58.73$^s$, Dec. (J2000) = +61° 47' 46.3"

Cross Identification: 2MASS J02535873+6147462; UCAC4 759-023133

Variability Type: Var (Unspecified)

Magnitude: Max. 12.4 V, Min. 12.5 V

Period: 0.27534(2) d

Epoch: 2456630.456(1) HJD

Ensemble Comparison Stars: UCAC4 760-023181 (APASS 11.403 V);
UCAC4 760-023174 (APASS 11.373 V)

Check Star: UCAC4 760-023206

Notes: B–V = 0.47; J–K = 0.54. Mean magnitude change superposed on the 0.27534 d. (amplitude 0.05 magnitude) variations. Type ELL with P = 0.5507 d. is possible. Check star stable with uniform 0.03-magnitude scattering. Fourier spectrum, light curve, and Phase plot are shown in Figures 31, 32, and 33.

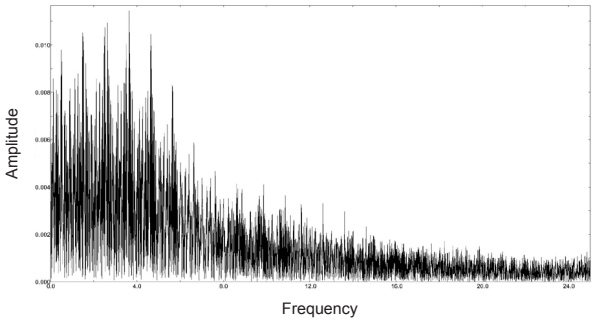

Figure 31. Fourier spectrum of GSC 04048-00441 (after detrending for the progressive fading).

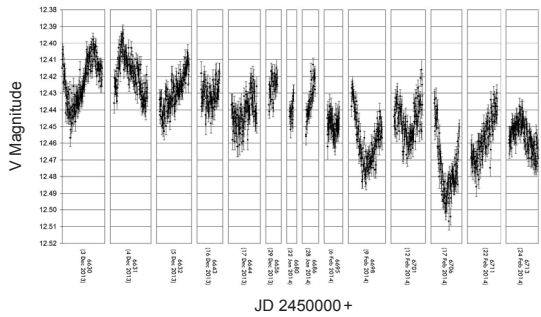

Figure 32. Light curve of GSC 04048-00441.

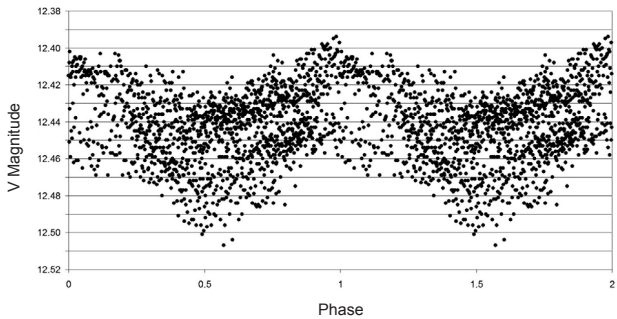

Figure 33. Phase plot of GSC 04048-00441; filled circles denote Furgoni data.

## 6. Acknowledgements

I wish to thank Sebastian Otero, member of the VSX team and AAVSO external consultant, for his helpful comments on the new variable stars discovered. This work has made use of the VIZIER catalogue access tool, CDS, Strasbourg, France, and the International Variable Star Index (VSX) operated by the AAVSO, Cambridge, Massachusetts. This work has made use of NASA's Astrophysics Data System and data products from the Two Micron All Sky Survey, which is a joint project of the University of Massachusetts and the Infrared Processing and Analysis Center/California Institute of Technology, funded by the National Aeronautics and Space Administration and the National Science Foundation.

This work has made use of the ASAS3 Public Catalog, NSVS data obtained from the Sky Database for Objects in Time-Domain operated by the Los Alamos National Laboratory, and data obtained from the SuperWASP Public Archive operated by the WASP consortium, which consists of representatives from the Queen's University Belfast, the University of Cambridge (Wide Field Astronomy Unit), Instituto de Astrofisica de Canarias, the Isaac Newton Group of Telescopes (La Palma), the University of Keele, the University of Leicester, the Open University, the University of St Andrews, and the South African Astronomical Observatory.

This work has made use of The Fourth U.S. Naval Observatory CCD Astrograph Catalog (UCAC4).